# Sedeonic theory of massless fields

V.L.Mironov, S.V.Mironov and S.A.Korolev

Institute for physics of microstructures RAS, 603950, Nizhniy Novgorod, GSP-105, Russia
e-mail: mironov@ipm.sci-nnov.ru

In present paper we develop the description of massless fields on the basis of space-time algebra of sixteen-component sedeons. The generalized sedeonic second-order equation for unified gravitoelectromagnetic (GE) field describing simultaneously gravity and electromagnetism is proposed. The second-order relations for the GE field energy, momentum and Lorentz invariants are derived. We consider also the generalized sedeonic first-order equation for the massless neutrino field. The second-order relations for the neutrino potentials analogues to the Pointing theorem and Lorentz invariant relations in gravitoelectromagnetism are also derived.

## 1. Introduction

The linearized weak field equations of general relativity [1, 2] can be represented as the set of Maxwell-like equations for the vectorial gravitoelectric and gravitomagnetic fields [3-5]. This linear approach (so-called "post-Newtonian approximation" or "gravitoelectromagnetism") is widely used in astrophysics for the analysis of interactions between moving and spinning masses (see for example Refs. 6-9). However, the vector algebra, which is usually used for the formulation of electromagnetism and gravitoelectromagnetism, does not adequately specify the space-time properties of gravitational and electromagnetic fields. Since the Maxwell-like equations for electromagnetic field (and corresponding linear equations for gravitational field) are the system of four equations for scalar, pseudoscalar, vector and pseudovector values the application of the eight-component algebras is more appropriate. There are different approaches based on algebra of hypercomplex numbers and Clifford algebras to formulate electromagnetism [10-17] and linear gravity [18, 19] taking into account above mentioned space symmetry. However, the consideration of total space-time field's symmetry requires the introduction of sixteen-component space-time values. It is known some attempts to develop a field theory on the basis of sixteen-component structures such as hypercomplex numbers sedenions [20-24] and hypercomplex multivectors generating associative space-time Clifford algebras [25,26]. However, these attempts have not made appreciable progress.

Recently we proposed the sixteen-component sedeonic space-time algebra [27,28] to modify the field theory equations. In present paper we develop the consideration of massless fields in terms of sedeonic potentials. We show that linear model of gravitoelectromagnetism enables the introduction of unified sedeonic gravitoelectromagnetic (GE) field describing simultaneously gravity and electromagnetism.

This paper has the following structure. In Sec. 2 we briefly review the basic properties of sedeonic algebra. In Sec. 3 the symmetric form of Newton-Coulomb low of gravitoelectromagnetism is discussed. Sec. 4 is devoted to the formulation of symmetric equations for unified gravitoelectromagnetic fields. In Sec. 5 and 6 the second-order relations for energy, momentum and Lorentz invariants of GE field are derived. In Sec. 7 we analyze the generalized sedeonic equation for the massless neutrino field. Finally, in Sec. 8 the second-order relations for neutrino potentials are discussed.

## 2. Sedeonic space-time algebra

To begin with we will briefly review the basic properties of sedeons. The sixteen-component sedeons take into account total space-time symmetry of physical fields. The sedeonic algebra encloses four groups of values, which are differed with respect to spatial and time inversion.

(1) Absolute scalars ($V$) and absolute vectors ($\vec{V}$) are not transformed under spatial and time inversion.
(2) Time scalars ($V_t$) and time vectors ($\vec{V}_t$) are changed (in sign) under time inversion and are not transformed under spatial inversion.
(3) Space scalars ($V_r$) and space vectors ($\vec{V}_r$) are changed under spatial inversion and are not transformed under time inversion.
(4) Space-time scalars ($V_{tr}$) and space-time vectors ($\vec{V}_{tr}$) are changed under spatial and time inversion.

The indexes **t** and **r** indicate the transformations (**t** for time inversion and **r** for spatial inversion), which change the corresponding values. All introduced values can be integrated into one space-time object sedeon $\tilde{V}$, which is defined by the following expression:

$$\tilde{V} = V + \vec{V} + V_t + \vec{V}_t + V_r + \vec{V}_r + V_{tr} + \vec{V}_{tr}. \tag{2.1}$$

Let us introduce scalar-vector basis $\mathbf{a_0}$, $\mathbf{a_1}$, $\mathbf{a_2}$, $\mathbf{a_3}$, where value $\mathbf{a_0} \equiv 1$ is absolute scalar unit and the values $\mathbf{a_1}$, $\mathbf{a_2}$, $\mathbf{a_3}$ are absolute unit vectors generating the right Cartesian basis. We introduce also four space-time scalar units $\mathbf{e_0}$, $\mathbf{e_1}$, $\mathbf{e_2}$, $\mathbf{e_3}$, where value $\mathbf{e_0} \equiv 1$ is a absolute scalar unit; $\mathbf{e_1} \equiv \mathbf{e_t}$ is a time scalar unit; $\mathbf{e_2} \equiv \mathbf{e_r}$ is a space scalar unit; $\mathbf{e_3} \equiv \mathbf{e_{tr}}$ is a space-time scalar unit. Using scalar-vector basis $\mathbf{a_k}$ (k = 0, 1, 2, 3) and space-time scalar units $\mathbf{e_n}$ (n = 0, 1, 2, 3) we can introduce unified sedeonic components $V_{nk}$ in accordance with the following relations

$$\begin{aligned}
V &= \mathbf{e_0} V_{00} \mathbf{a_0}, \\
\vec{V} &= \mathbf{e_0} \left( V_{01} \mathbf{a_1} + V_{02} \mathbf{a_2} + V_{03} \mathbf{a_3} \right), \\
V_t &= \mathbf{e_1} V_{10} \mathbf{a_0}, \\
\vec{V}_t &= \mathbf{e_1} \left( V_{11} \mathbf{a_1} + V_{12} \mathbf{a_2} + V_{13} \mathbf{a_3} \right), \\
V_r &= \mathbf{e_2} V_{20} \mathbf{a_0}, \\
\vec{V}_r &= \mathbf{e_2} \left( V_{21} \mathbf{a_1} + V_{22} \mathbf{a_2} + V_{23} \mathbf{a_3} \right), \\
V_{rt} &= \mathbf{e_3} V_{30} \mathbf{a_0}, \\
\vec{V}_{rt} &= \mathbf{e_3} \left( V_{31} \mathbf{a_1} + V_{32} \mathbf{a_2} + V_{33} \mathbf{a_3} \right).
\end{aligned} \tag{2.2}$$

Then sedeon (2.1) can be written in the following expanded form:

$$\begin{aligned}
\tilde{V} = &\mathbf{e_0} \left( V_{00} \mathbf{a_0} + V_{01} \mathbf{a_1} + V_{02} \mathbf{a_2} + V_{03} \mathbf{a_3} \right) + \mathbf{e_1} \left( V_{10} \mathbf{a_0} + V_{11} \mathbf{a_1} + V_{12} \mathbf{a_2} + V_{13} \mathbf{a_3} \right) \\
&+ \mathbf{e_2} \left( V_{20} \mathbf{a_0} + V_{21} \mathbf{a_1} + V_{22} \mathbf{a_2} + V_{23} \mathbf{a_3} \right) + \mathbf{e_3} \left( V_{30} \mathbf{a_0} + V_{31} \mathbf{a_1} + V_{32} \mathbf{a_2} + V_{33} \mathbf{a_3} \right).
\end{aligned} \tag{2.3}$$

The sedeonic components $V_{nk}$ are numbers (complex in general). Further we will use symbol 1 instead units $\mathbf{a_0}$ and $\mathbf{e_0}$ for simplicity.

The multiplication and commutation rules for sedeonic absolute unit vectors $\mathbf{a_1}, \mathbf{a_2}, \mathbf{a_3}$ and space-time units $\mathbf{e_1}, \mathbf{e_2}, \mathbf{e_3}$ are presented in tables 1 and 2.

*Table 1.*

|     | $\mathbf{a_1}$ | $\mathbf{a_2}$ | $\mathbf{a_3}$ |
|-----|-----|-----|-----|
| $\mathbf{a_1}$ | 1 | $i\mathbf{a_3}$ | $-i\mathbf{a_2}$ |
| $\mathbf{a_2}$ | $-i\mathbf{a_3}$ | 1 | $i\mathbf{a_1}$ |
| $\mathbf{a_3}$ | $i\mathbf{a_2}$ | $-i\mathbf{a_1}$ | 1 |

*Table 2.*

|     | $\mathbf{e_1}$ | $\mathbf{e_2}$ | $\mathbf{e_3}$ |
|-----|-----|-----|-----|
| $\mathbf{e_1}$ | 1 | $i\mathbf{e_3}$ | $-i\mathbf{e_2}$ |
| $\mathbf{e_2}$ | $-i\mathbf{e_3}$ | 1 | $i\mathbf{e_1}$ |
| $\mathbf{e_3}$ | $i\mathbf{e_2}$ | $-i\mathbf{e_1}$ | 1 |

In the tables and further the value $i$ is the imaginary unit ($i^2 = -1$). Note that sedeonic units $\mathbf{e_1}$, $\mathbf{e_2}$, $\mathbf{e_3}$ and unit vectors $\mathbf{a_1}$, $\mathbf{a_2}$, $\mathbf{a_3}$ generate the anticommutative algebras, but $\mathbf{e_1}, \mathbf{e_2}, \mathbf{e_3}$ commute with $\mathbf{a_1}, \mathbf{a_2}, \mathbf{a_3}$. Thus the sedeon $\tilde{V}$ is the complicated space-time object consisting of absolute scalar, space scalar, time scalar, space-time scalar, absolute vector, space vector, time vector and space-time vector.

Introducing designations of space-time sedeon-scalars

$$V_0 = (V_{00} + \mathbf{e_1}V_{10} + \mathbf{e_2}V_{20} + \mathbf{e_3}V_{30}),$$
$$V_1 = (V_{01} + \mathbf{e_1}V_{11} + \mathbf{e_2}V_{21} + \mathbf{e_3}V_{31}),$$
$$V_2 = (V_{02} + \mathbf{e_1}V_{12} + \mathbf{e_2}V_{22} + \mathbf{e_3}V_{32}), \qquad (2.4)$$
$$V_3 = (V_{03} + \mathbf{e_1}V_{13} + \mathbf{e_2}V_{23} + \mathbf{e_3}V_{33}),$$

we can write the sedeon (2.3) in the compact form

$$\tilde{V} = V_0 + V_1\mathbf{a_1} + V_2\mathbf{a_2} + V_3\mathbf{a_3}, \qquad (2.5)$$

or introducing the sedeon-vector

$$\vec{V} = \vec{V} + \vec{V}_t + \vec{V}_r + \vec{V}_{tr} = V_1\mathbf{a_1} + V_2\mathbf{a_2} + V_3\mathbf{a_3}, \qquad (2.6)$$

it can be represented in following compact form

$$\tilde{V} = V_0 + \vec{V}. \qquad (2.7)$$

Further we will indicate sedeon-scalars and sedeon-vectors with the bold capital letters.

Let us consider the sedeonic multiplication in detail. The sedeonic product of two sedeons $\tilde{A}$ and $\tilde{B}$ can be represented in the following form

$$\tilde{A}\tilde{B} = \left(A_0 + \vec{A}\right)\left(B_0 + \vec{B}\right) = A_0B_0 + A_0\vec{B} + \vec{A}B_0 + \left(\vec{A}\cdot\vec{B}\right) + \left[\vec{A}\times\vec{B}\right]. \qquad (2.8)$$

Here we denoted the sedeonic scalar multiplication of two sedeon-vectors (internal product) by symbol "$\cdot$" and round brackets

$$\left(\vec{A}\cdot\vec{B}\right) = A_1B_1 + A_2B_2 + A_3B_3, \qquad (2.9)$$

and sedeonic vector multiplication (external product) by symbol "$\times$" and square brackets,

$$\left[\vec{A}\times\vec{B}\right] = i(A_2B_3 - A_3B_2)\mathbf{a_1} + i(A_3B_1 - A_1B_3)\mathbf{a_2} + i(A_1B_2 - A_2B_1)\mathbf{a_3}. \qquad (2.10)$$

In (2.9) and (2.10) the multiplication of sedeonic components is performed in accordance with (2.6) and table 2. Note that in sedeonic algebra the vector triple product has some difference from Gibbs vector algebra. Let us consider three absolute vectors $\vec{A}$, $\vec{B}$ and $\vec{C}$. Then the formula for the vector triple product in sedeonic algebra has the following form:

$$\left[\vec{A}\times\left[\vec{B}\times\vec{C}\right]\right]=-\vec{B}\left(\vec{A}\cdot\vec{C}\right)+\vec{C}\left(\vec{A}\cdot\vec{B}\right). \tag{2.11}$$

In the next sections we apply the sedeonic algebra to the unified description of gravitational and electromagnetic fields.

## 3. Newton - Coulomb low

It is known that Coulomb's law for the force of electrical interaction between two charged point bodies is written as follows:

$$\vec{F}_{e12}=\frac{q_{e1}q_{e2}}{r_{12}^3}\vec{r}_{12}, \tag{3.1}$$

where $q_{e1}$ and $q_{e2}$ are electrical charges, $\vec{r}_{12}$ is a vector directed from body 1 to body 2, $r_{12}$ is the separation between point bodies, which is equal to modulus of $\vec{r}_{12}$. For a symmetric description of electromagnetic and gravitational phenomena, we introduce the gravitational charge $q_g$, considered previously in Refs 7 and 29:

$$q_g=\sqrt{G}m, \tag{3.2}$$

where $G$ is the gravitational constant, $m$ is a mass of gravitating body. Then Newton's law for gravitational force between two point bodies can be written in the form of Coulomb's law:

$$\vec{F}_{g12}=-\frac{q_{g1}q_{g2}}{r_{12}^3}\vec{r}_{12}. \tag{3.3}$$

Simultaneous consideration of gravitational and electromagnetic fields leads us to another symmetry connected with charge conjugation. The idea consists in an introduction of additional noncommutative and nonassociative units associated with electrical and gravitational charges. Let us introduce two new units. First one is electrical unit $\varepsilon_1 \equiv \varepsilon_e$, which is changed (in sign) under electrical charge conjugation. Second one is gravitational unit $\varepsilon_2 \equiv \varepsilon_g$, which is changed (in sign) under gravitational charge conjugation. We emphasize that these units are noncommutative and nonassociative. For example:

$$\begin{aligned}\varepsilon_1\varepsilon_2&=-\varepsilon_2\varepsilon_1,\\ \varepsilon_1\varepsilon_2\varepsilon_1&=-\varepsilon_2.\end{aligned} \tag{3.4}$$

We assume that in the classical gravitoelectrodynamics there is no direct interaction between gravitational and electrical charges. Then the rules of multiplication for units $\varepsilon_1$ and $\varepsilon_1$ should be chosen as follows:

*Table 3. Multiplication rules for $\varepsilon_1$ and $\varepsilon_1$ units.*

|  | $\varepsilon_1$ | $\varepsilon_2$ |
| --- | --- | --- |
| $\varepsilon_1$ | 1 | 0 |
| $\varepsilon_2$ | 0 | 1 |

Following this approach, the generalized gravitoelectromagnetic charge $Q$ can be presented as

$$Q = \varepsilon_1 q_e - i\varepsilon_2 q_g. \tag{3.5}$$

Then the operations of electrical charge conjugation ($\hat{I}_e$), gravitational charge conjugation ($\hat{I}_g$), and electrogravitational charge conjugation ($\hat{I}_{eg}$) can be presented as

$$\hat{I}_e Q = \varepsilon_2 Q \varepsilon_2 = -\varepsilon_1 q_e - i\varepsilon_2 q_g, \tag{3.6}$$

$$\hat{I}_g Q = \varepsilon_1 Q \varepsilon_1 = \varepsilon_1 q_e + i\varepsilon_2 q_g, \tag{3.7}$$

$$\hat{I}_{eg} Q = \varepsilon_2 \varepsilon_1 Q \varepsilon_1 \varepsilon_2 = -\varepsilon_1 q_e + i\varepsilon_2 q_g. \tag{3.8}$$

The generalized Newton - Coulomb law can be written as:

$$\vec{F}_{12} = \frac{Q_1 Q_2}{r_{12}^3} \vec{r}_{12}. \tag{3.9}$$

Indeed, relation (3.9) gives us correct expression for the generalized force between two massive electrically charged point bodies

$$\vec{F}_{12} = \frac{q_{e1} q_{e2}}{r_{12}^3} \vec{r}_{12} - \frac{q_{g1} q_{g2}}{r_{12}^3} \vec{r}_{12}. \tag{3.10}$$

Analogously one can introduce the generalized GE field $\vec{E}$ as

$$\vec{E} = \varepsilon_1 \vec{E}_e + i\varepsilon_2 \vec{E}_g, \tag{3.11}$$

where $\vec{E}_e$ is electrical field intensity, $\vec{E}_g$ is gravitational field intensity. Then the generalized Newton - Coulomb law can be written in the following form:

$$\vec{F}_{12} = \vec{E}_1 Q_2. \tag{3.12}$$

Indeed, this relation leads us to the correct expression for the generalized GE force

$$\vec{F}_{12} = \vec{E}_{e1} q_{e2} + \vec{E}_{g1} q_{g2}. \tag{3.13}$$

## 4. Generalized sedeonic equations for gravitoelectromagnetic field

The sedeonic formalism enables the representation of gravitational and electromagnetic fields as one uniformed gravitoelectromagnetic field. Indeed, the generalized sedeonic second-order equation for massless fields can be presented in the following form [27]:

$$\left( i\mathbf{e_t} \frac{1}{c} \frac{\partial}{\partial t} - \mathbf{e_r} \vec{\nabla} \right) \left( i\mathbf{e_t} \frac{1}{c} \frac{\partial}{\partial t} - \mathbf{e_r} \vec{\nabla} \right) \tilde{W} = \tilde{J}. \tag{4.1}$$

Let us consider the potential of GE field as:

$$\tilde{W} = i\mathbf{e_t} \varphi_e + \mathbf{e_r} \vec{A}_e + i\left( i\mathbf{e_t} \varphi_g + \mathbf{e_r} \vec{A}_g \right), \tag{4.2}$$

where $\varphi_e, \vec{A}_e, \varphi_g$ and $\vec{A}_g$, are real scalar and vector potentials of electromagnetic (index $e$) and gravitational (index $g$) fields. Hereafter we mean that electrical values contain $\varepsilon_1$ and gravitational values contain $\varepsilon_2$ units, but we omit them to simplify the farther expressions.

Let us also consider the generalized sedeonic source of GE field

$$\tilde{J} = -4\pi\left(i\mathbf{e_t}\rho_e + \mathbf{e_r}\frac{1}{c}\vec{j}_e\right) + 4\pi i\left(i\mathbf{e_t}\rho_g + \mathbf{e_r}\frac{1}{c}\vec{j}_g\right), \quad (4.3)$$

where $\rho_e$ is a volume density of electrical charge; $\vec{j}_e$ is a volume density of electrical current; $\rho_g$ is a volume density of gravity charge; $\vec{j}_g$ is a volume density of gravitational current [7]. Here we do not consider the magnetic and gravitomagnetic monopoles and corresponding magnetic currents. Then the sedeonic GE field equation (3.1) can be represented in the following form:

$$\left(i\mathbf{e_t}\frac{1}{c}\frac{\partial}{\partial t} - \mathbf{e_r}\vec{\nabla}\right)\left(i\mathbf{e_t}\frac{1}{c}\frac{\partial}{\partial t} - \mathbf{e_r}\vec{\nabla}\right)\left(i\mathbf{e_t}\varphi_e + \mathbf{e_r}\vec{A}_e + i\left(i\mathbf{e_t}\varphi_g + \mathbf{e_r}\vec{A}_g\right)\right) =$$
$$= -4\pi\left(i\mathbf{e_t}\rho_e + \mathbf{e_r}\frac{1}{c}\vec{j}_e\right) + 4\pi i\left(i\mathbf{e_t}\rho_g + \mathbf{e_r}\frac{1}{c}\vec{j}_g\right). \quad (4.4)$$

This equation describes simultaneously electromagnetic and gravitational fields. Performing sedeonic multiplication of operators in the left part and separating real and imaginary parts (or terms with $\varepsilon_1$ and $\varepsilon_2$ units) in time and space scalars and vectors we get the system of wave equations for the components of GE potential

$$\left(\frac{1}{c}\frac{\partial^2}{\partial t^2} - \Delta\right)\varphi_e = 4\pi\rho_e, \quad (4.5)$$

$$\left(\frac{1}{c}\frac{\partial^2}{\partial t^2} - \Delta\right)\vec{A}_e = 4\pi\frac{1}{c}\vec{j}_e, \quad (4.6)$$

$$\left(\frac{1}{c}\frac{\partial^2}{\partial t^2} - \Delta\right)\varphi_g = -4\pi\rho_g, \quad (4.7)$$

$$\left(\frac{1}{c}\frac{\partial^2}{\partial t^2} - \Delta\right)\vec{A}_g = -4\pi\frac{1}{c}\vec{j}_g. \quad (4.8)$$

On the other hand, equation (4.4) can be represented as the system of first-order Maxwell equation for electromagnetic and gravitational fields. Let us consider the sequential action of operators in Eq. (4.4). After the action of first operator we obtain

$$\left(i\mathbf{e_t}\frac{1}{c}\frac{\partial}{\partial t} - \mathbf{e_r}\vec{\nabla}\right)\left(i\mathbf{e_t}\varphi_e + \mathbf{e_r}\vec{A}_e + i\left(i\mathbf{e_t}\varphi_g + \mathbf{e_r}\vec{A}_g\right)\right) =$$
$$= -\frac{1}{c}\frac{\partial\varphi_e}{\partial t} - \mathbf{e_{rt}}\frac{1}{c}\frac{\partial\vec{A}_e}{\partial t} - \mathbf{e_{rt}}\vec{\nabla}\varphi_e - \left(\vec{\nabla}\cdot\vec{A}_e\right) - \left[\vec{\nabla}\times\vec{A}_e\right] \quad (4.9)$$
$$+i\left\{-\frac{1}{c}\frac{\partial\varphi_g}{\partial t} - \mathbf{e_{rt}}\frac{1}{c}\frac{\partial\vec{A}_g}{\partial t} - \mathbf{e_{rt}}\vec{\nabla}\varphi_g - \left(\vec{\nabla}\cdot\vec{A}_g\right) - \left[\vec{\nabla}\times\vec{A}_g\right]\right\}.$$

This expression enables the introduction of scalar and vector intensities of GE field in the following form:

$$f_e = -\frac{1}{c}\frac{\partial \varphi_e}{\partial t} - (\vec{\nabla} \cdot \vec{A}_e),$$

$$\vec{E}_e = -\vec{\nabla}\varphi_e - \frac{1}{c}\frac{\partial \vec{A}_e}{\partial t},$$

$$\vec{H}_e = -i\left[\vec{\nabla} \times \vec{A}_e\right],$$

$$f_g = -\frac{1}{c}\frac{\partial \varphi_g}{\partial t} - (\vec{\nabla} \cdot \vec{A}_g),$$

$$\vec{E}_g = -\vec{\nabla}\varphi_g - \frac{1}{c}\frac{\partial \vec{A}_g}{\partial t},$$

$$\vec{H}_g = -i\left[\vec{\nabla} \times \vec{A}_g\right].$$
(4.10)

Using the definitions (4.10), expression (4.9) can be rewritten as

$$\left(i\mathbf{e_t}\frac{1}{c}\frac{\partial}{\partial t} - \mathbf{e_r}\vec{\nabla}\right)\left(i\mathbf{e_t}\varphi_e + \mathbf{e_r}\vec{A}_e + i\left(i\mathbf{e_t}\varphi_g + \mathbf{e_r}\vec{A}_g\right)\right)$$
$$= f_e + \mathbf{e_{rt}}\vec{E}_e - i\vec{H}_e + i\left(f_g + \mathbf{e_{rt}}\vec{E}_g - i\vec{H}_g\right).$$
(4.11)

Then the generalized equation (4.4) can be represented in the following form:

$$\left(i\mathbf{e_t}\frac{1}{c}\frac{\partial}{\partial t} - \mathbf{e_r}\vec{\nabla}\right)\left(f_e + \mathbf{e_{rt}}\vec{E}_e - i\vec{H}_e + i\left(f_g + \mathbf{e_{rt}}\vec{E}_g - i\vec{H}_g\right)\right)$$
$$= -4\pi\left(i\mathbf{e_t}\rho_e + \mathbf{e_r}\frac{1}{c}\vec{j}_e\right) + 4\pi i\left(i\mathbf{e_t}\rho_g + \mathbf{e_r}\frac{1}{c}\vec{j}_g\right).$$
(4.12)

Applying operator $\left(i\mathbf{e_t}\frac{1}{c}\frac{\partial}{\partial t} - \mathbf{e_r}\vec{\nabla}\right)$ to both parts of expression (4.12) one can obtain the second-order wave equations for the field intensities in the following form:

$$\left(\frac{1}{c^2}\frac{\partial^2}{\partial t^2} - \Delta\right)f_e = -\frac{4\pi}{c}\left\{\frac{\partial \rho_e}{\partial t} + (\vec{\nabla} \cdot \vec{j}_e)\right\},$$
(4.13)

$$\left(\frac{1}{c^2}\frac{\partial^2}{\partial t^2} - \Delta\right)\vec{E}_e = -4\pi\vec{\nabla}\rho_e - \frac{4\pi}{c^2}\frac{\partial \vec{j}_e}{\partial t},$$
(4.14)

$$\left(\frac{1}{c^2}\frac{\partial^2}{\partial t^2} - \Delta\right)\vec{H}_e = -i\frac{4\pi}{c}\left[\vec{\nabla} \times \vec{j}_e\right],$$
(4.15)

$$\left(\frac{1}{c^2}\frac{\partial^2}{\partial t^2} - \Delta\right)f_g = \frac{4\pi}{c}\left\{\frac{\partial \rho_g}{\partial t} + (\vec{\nabla} \cdot \vec{j}_g)\right\},$$
(4.16)

$$\left(\frac{1}{c^2}\frac{\partial^2}{\partial t^2} - \Delta\right)\vec{E}_g = 4\pi\vec{\nabla}\rho_g + \frac{4\pi}{c^2}\frac{\partial \vec{j}_g}{\partial t},$$
(4.17)

$$\left(\frac{1}{c^2}\frac{\partial^2}{\partial t^2} - \Delta\right)\vec{H}_g = i\frac{4\pi}{c}\left[\vec{\nabla} \times \vec{j}_g\right].$$
(4.18)

In the absence of processes of matter nucleation and annihilation, we can assume the following conservation laws

$$\frac{\partial \rho_e}{\partial t} + \left(\vec{\nabla} \cdot \vec{j}_e\right) = 0, \tag{4.19}$$

$$\frac{\partial \rho_g}{\partial t} + \left(\vec{\nabla} \cdot \vec{j}_g\right) = 0, \tag{4.20}$$

and can take the scalar fields $f_e$ and $f_g$ equal to zero. This is equivalent to the Lorentz gauge conditions (see expressions 4.10):

$$f_e = \frac{1}{c}\frac{\partial \varphi_e}{\partial t} + (\vec{\nabla} \cdot \vec{A}_e) = 0, \tag{4.21}$$

$$f_g = \frac{1}{c}\frac{\partial \varphi_g}{\partial t} + (\vec{\nabla} \cdot \vec{A}_g) = 0. \tag{4.22}$$

In this case, the expression (4.11) is rewritten as

$$\left(i\mathbf{e_t}\frac{1}{c}\frac{\partial}{\partial t} - \mathbf{e_r}\vec{\nabla}\right)\left(i\mathbf{e_t}\varphi_e + \mathbf{e_r}\vec{A}_e + i\left(i\mathbf{e_t}\varphi_g + \mathbf{e_r}\vec{A}_g\right)\right) = \mathbf{e_{rt}}\vec{E}_e - i\vec{H}_e + i\left(\mathbf{e_{rt}}\vec{E}_g - i\vec{H}_g\right), \tag{4.21}$$

and equation (4.12) can be represented as

$$\left(i\mathbf{e_t}\frac{1}{c}\frac{\partial}{\partial t} - \mathbf{e_r}\vec{\nabla}\right)\left(\mathbf{e_{rt}}\vec{E}_e - i\vec{H}_e + i\left(\mathbf{e_{rt}}\vec{E}_g - i\vec{H}_g\right)\right)$$
$$= -4\pi\left(i\mathbf{e_t}\rho_e + \mathbf{e_r}\frac{1}{c}\vec{j}_e\right) + 4\pi i\left(i\mathbf{e_t}\rho_g + \mathbf{e_r}\frac{1}{c}\vec{j}_g\right). \tag{4.22}$$

Performing sedeonic multiplication in the left part of equation (4.22) we get:

$$\left(-i\mathbf{e_t}\left(\vec{\nabla}\cdot\vec{E}_e\right) + i\mathbf{e_r}\left(\vec{\nabla}\cdot\vec{H}_e\right) - i\mathbf{e_t}\left[\vec{\nabla}\times\vec{E}_e\right] + \mathbf{e_t}\frac{1}{c}\frac{\partial \vec{H}_e}{\partial t} + i\mathbf{e_r}\left[\vec{\nabla}\times\vec{H}_e\right] + \mathbf{e_r}\frac{1}{c}\frac{\partial \vec{E}_e}{\partial t}\right)$$
$$+ i\left(-i\mathbf{e_t}\left(\vec{\nabla}\cdot\vec{E}_g\right) + i\mathbf{e_r}\left(\vec{\nabla}\cdot\vec{H}_g\right) - i\mathbf{e_t}\left[\vec{\nabla}\times\vec{E}_g\right] + \mathbf{e_t}\frac{1}{c}\frac{\partial \vec{H}_g}{\partial t} + i\mathbf{e_r}\left[\vec{\nabla}\times\vec{H}_g\right] + \mathbf{e_r}\frac{1}{c}\frac{\partial \vec{E}_g}{\partial t}\right) \tag{4.23}$$
$$= -4\pi\left(i\mathbf{e_t}\rho_e + \mathbf{e_r}\frac{1}{c}\vec{j}_e\right) + 4\pi i\left(i\mathbf{e_t}\rho_g + \mathbf{e_r}\frac{1}{c}\vec{j}_g\right).$$

Separating terms with different space-time properties, we get a system of Maxwell's equations for the GE field

$$-i\left[\vec{\nabla}\times\left(\vec{E}_e + i\vec{E}_g\right)\right] = -\frac{1}{c}\frac{\partial}{\partial t}\left(\vec{H}_e + i\vec{H}_g\right),$$
$$-i\left[\vec{\nabla}\times\left(\vec{H}_e + i\vec{H}_g\right)\right] = \frac{4\pi}{c}\left(\vec{j}_e - i\vec{j}_g\right) + \frac{1}{c}\frac{\partial}{\partial t}\left(\vec{E}_e + i\vec{E}_g\right),$$
$$\left(\vec{\nabla}\cdot\left(\vec{E}_e + i\vec{E}_g\right)\right) = 4\pi\left(\rho_e - i\rho_g\right), \tag{4.24}$$
$$\left(\vec{\nabla}\cdot\left(\vec{H}_e + i\vec{H}_g\right)\right) = 0.$$

Separating real and imaginary parts (or terms with $\varepsilon_1$ and $\varepsilon_2$ units), we obtain two systems of Maxwell equations for electromagnetic and gravitational fields. For the electromagnetic field we get following system:

$$-i\left[\vec{\nabla}\times\vec{E}_e\right] = -\frac{1}{c}\frac{\partial \vec{H}_e}{\partial t},$$

$$-i\left[\vec{\nabla}\times\vec{H}_e\right] = \frac{4\pi}{c}\vec{j}_e + \frac{1}{c}\frac{\partial \vec{E}_e}{\partial t},\qquad(4.25)$$

$$\left(\vec{\nabla}\cdot\vec{E}_e\right) = 4\pi\rho_e,$$

$$\left(\vec{\nabla}\cdot\vec{H}_e\right) = 0.$$

On the other hand, for the gravitational field we obtain

$$-i\left[\vec{\nabla}\times\vec{E}_g\right] = -\frac{1}{c}\frac{\partial \vec{H}_g}{\partial t},$$

$$-i\left[\vec{\nabla}\times\vec{H}_g\right] = -\frac{4\pi}{c}\vec{j}_g + \frac{1}{c}\frac{\partial \vec{E}_g}{\partial t},\qquad(4.26)$$

$$\left(\vec{\nabla}\cdot\vec{E}_g\right) = -4\pi\rho_g,$$

$$\left(\vec{\nabla}\cdot\vec{H}_g\right) = 0.$$

Thus, we have shown that the generalized equation (4.4) correctly describes the unified GE field.

## 5. Sedeonic relations for energy and momentum of GE field

The sedeonic algebra allows one to provide the combined calculus with electromagnetic and gravitational fields simultaneously. Multiplying the expression (4.22) on the sedeon $\mathbf{e_{rt}}\vec{E}_e - i\vec{H}_e + i\left(\mathbf{e_{rt}}\vec{E}_g - i\vec{H}_g\right)$ from the left we obtain

$$\left(\mathbf{e_{rt}}\vec{E}_e - i\vec{H}_e + i\left(\mathbf{e_{rt}}\vec{E}_g - i\vec{H}_g\right)\right)\left(i\mathbf{e_t}\frac{1}{c}\frac{\partial}{\partial t} - \mathbf{e_r}\vec{\nabla}\right)\left(\mathbf{e_{rt}}\vec{E}_e - i\vec{H}_e + i\left(\mathbf{e_{rt}}\vec{E}_g - i\vec{H}_g\right)\right)$$

$$= -4\pi\left(\mathbf{e_{rt}}\vec{E}_e - i\vec{H}_e + i\left(\mathbf{e_{rt}}\vec{E}_g - i\vec{H}_g\right)\right)\left(i\mathbf{e_t}\rho_e + \mathbf{e_r}\frac{1}{c}\vec{j}_e - i\left(i\mathbf{e_t}\rho_g + \mathbf{e_r}\frac{1}{c}\vec{j}_g\right)\right).\qquad(5.1)$$

Then performing sedeonic multiplication, we get the following expression:

$$-i\mathbf{e_t}\left\{\frac{1}{2c}\frac{\partial}{\partial t}\left\{\left(\vec{E}_e + i\vec{E}_g\right)^2 + \left(\vec{H}_e + i\vec{H}_g\right)^2\right\} - i\left(\vec{\nabla}\cdot\left[\left(\vec{E}_e + i\vec{E}_g\right)\times\left(\vec{H}_e + i\vec{H}_g\right)\right]\right)\right\}$$

$$+\mathbf{e_r}\left\{i\frac{1}{c}\left(\left(\vec{E}_e + i\vec{E}_g\right)\cdot\frac{\partial\left(\vec{H}_e + i\vec{H}_g\right)}{\partial t}\right) - i\frac{1}{c}\left(\left(\vec{H}_e + i\vec{H}_g\right)\cdot\frac{\partial\left(\vec{E}_e + i\vec{E}_g\right)}{\partial t}\right)\right.$$

$$\left.+\left(\left(\vec{E}_e + i\vec{E}_g\right)\cdot\left[\vec{\nabla}\times\left(\vec{E}_e + i\vec{E}_g\right)\right]\right) + \left(\left(\vec{H}_e + i\vec{H}_g\right)\cdot\left[\vec{\nabla}\times\left(\vec{H}_e + i\vec{H}_g\right)\right]\right)\right\}$$

$$+\mathbf{e_t}\left\{-i\frac{1}{c}\left[\left(\vec{E}_e + i\vec{E}_g\right)\times\frac{\partial\left(\vec{E}_e + i\vec{E}_g\right)}{\partial t}\right] - i\frac{1}{c}\left[\left(\vec{H}_e + i\vec{H}_g\right)\times\frac{\partial\left(\vec{H}_e + i\vec{H}_g\right)}{\partial t}\right]\right.$$

$$+\left(\vec{E}_e + i\vec{E}_g\right)\left(\vec{\nabla}\cdot\left(\vec{H}_e + i\vec{H}_g\right)\right) - \left(\vec{H}_e + i\vec{H}_g\right)\left(\vec{\nabla}\cdot\left(\vec{E}_e + i\vec{E}_g\right)\right)\qquad(5.2)$$

$$\left.+\left[\left(\vec{E}_e + i\vec{E}_g\right)\times\left[\vec{\nabla}\times\left(\vec{H}_e + i\vec{H}_g\right)\right]\right] - \left[\left(\vec{H}_e + i\vec{H}_g\right)\times\left[\vec{\nabla}\times\left(\vec{E}_e + i\vec{E}_g\right)\right]\right]\right\}$$

$$+\mathbf{e_r}\left\{i\frac{1}{c}\frac{\partial}{\partial t}\left[\left(\vec{E}_e+i\vec{E}_g\right)\times\left(\vec{H}_e+i\vec{H}_g\right)\right]-\frac{1}{2}\vec{\nabla}\left\{\left(\vec{E}_e+i\vec{E}_g\right)^2+\left(\vec{H}_e+i\vec{H}_g\right)^2\right\}\right.$$

$$\left.+\left(\vec{\nabla}\cdot\left(\vec{E}_e+i\vec{E}_g\right)\right)\left(\vec{E}_e+i\vec{E}_g\right)+\left(\vec{\nabla}\cdot\left(\vec{H}_e+i\vec{H}_g\right)\right)\left(\vec{H}_e+i\vec{H}_g\right)\right\}$$

$$=i\mathbf{e_t}\frac{4\pi}{c}\left(\left(\vec{E}_e+i\vec{E}_g\right)\cdot\left(\vec{j}_e-i\vec{j}_g\right)\right)+i\mathbf{e_r}\frac{4\pi}{c}\left(\left(\vec{H}_e+i\vec{H}_g\right)\cdot\left(\vec{j}_e-i\vec{j}_g\right)\right)$$

$$-4\pi\mathbf{e_t}\left\{\left(\rho_e-i\rho_g\right)\left(\vec{H}_e+i\vec{H}_g\right)-i\frac{1}{c}\left[\left(\vec{E}_e+i\vec{E}_g\right)\times\left(\vec{j}_e-i\vec{j}_g\right)\right]\right\}$$

$$+4\pi\mathbf{e_r}\left\{\left(\rho_e-i\rho_g\right)\left(\vec{E}_e+i\vec{E}_g\right)+i\frac{1}{c}\left[\left(\vec{H}_e+i\vec{H}_g\right)\times\left(\vec{j}_e-i\vec{j}_g\right)\right]\right\}.$$

Note that in this expression and further the operator $\vec{\nabla}$ acts on all right expression. For example, for any vectors $\vec{A}$ and $\vec{B}$ we have

$$\left(\vec{\nabla}\cdot\vec{A}\right)\vec{B}=\vec{B}\left(\vec{\nabla}\cdot\vec{A}\right)+\left(\vec{A}\cdot\vec{\nabla}\right)\vec{B}. \tag{5.3}$$

Equating the components with different space-time properties we get the following equations for the GE field intensities:

$$\frac{1}{8\pi}\frac{\partial}{\partial t}\left\{\left(\vec{E}_e+i\vec{E}_g\right)^2+\left(\vec{H}_e+i\vec{H}_g\right)^2\right\}-i\frac{c}{4\pi}\left(\vec{\nabla}\cdot\left[\left(\vec{E}_e+i\vec{E}_g\right)\times\left(\vec{H}_e+i\vec{H}_g\right)\right]\right)$$
$$+\left(\left(\vec{E}_e+i\vec{E}_g\right)\cdot\left(\vec{j}_e-i\vec{j}_g\right)\right)=0, \tag{5.4}$$

$$\frac{1}{4\pi}\left\{\left(\vec{E}_e+i\vec{E}_g\right)\cdot\frac{\partial\left(\vec{H}_e+i\vec{H}_g\right)}{\partial t}-\left(\vec{H}_e+i\vec{H}_g\right)\cdot\frac{\partial\left(\vec{E}_e+i\vec{E}_g\right)}{\partial t}\right\}$$
$$-i\frac{c}{4\pi}\left\{\left(\left(\vec{E}_e+i\vec{E}_g\right)\cdot\left[\vec{\nabla}\times\left(\vec{E}_e+i\vec{E}_g\right)\right]\right)+\left(\left(\vec{H}_e+i\vec{H}_g\right)\cdot\left[\vec{\nabla}\times\left(\vec{H}_e+i\vec{H}_g\right)\right]\right)\right\} \tag{5.5}$$
$$-\left(\left(\vec{H}_e+i\vec{H}_g\right)\cdot\left(\vec{j}_e-i\vec{j}_g\right)\right)=0,$$

$$-i\frac{1}{4\pi}\left\{\left[\left(\vec{E}_e+i\vec{E}_g\right)\times\frac{\partial\left(\vec{E}_e+i\vec{E}_g\right)}{\partial t}\right]+\left[\left(\vec{H}_e+i\vec{H}_g\right)\times\frac{\partial\left(\vec{H}_e+i\vec{H}_g\right)}{\partial t}\right]\right\}$$
$$+\frac{c}{4\pi}\left\{\left(\vec{E}_e+i\vec{E}_g\right)\left(\vec{\nabla}\cdot\left(\vec{H}_e+i\vec{H}_g\right)\right)-\left(\vec{H}_e+i\vec{H}_g\right)\left(\vec{\nabla}\cdot\left(\vec{E}_e+i\vec{E}_g\right)\right)\right\} \tag{5.6}$$
$$+\frac{c}{4\pi}\left\{\left[\left(\vec{E}_e+i\vec{E}_g\right)\times\left[\vec{\nabla}\times\left(\vec{H}_e+i\vec{H}_g\right)\right]\right]-\left[\left(\vec{H}_e+i\vec{H}_g\right)\times\left[\vec{\nabla}\times\left(\vec{E}_e+i\vec{E}_g\right)\right]\right]\right\}$$
$$+c\left(\vec{H}_e+i\vec{H}_g\right)\left(\rho_e-i\rho_g\right)-i\left[\left(\vec{E}_e+i\vec{E}_g\right)\times\left(\vec{j}_e-i\vec{j}_g\right)\right]=0,$$

$$-i\frac{1}{4\pi}\frac{\partial}{\partial t}\left[\left(\vec{E}_e+i\vec{E}_g\right)\times\left(\vec{H}_e+i\vec{H}_g\right)\right]+\frac{c}{8\pi}\vec{\nabla}\left\{\left(\vec{E}_e+i\vec{E}_g\right)^2+\left(\vec{H}_e+i\vec{H}_g\right)^2\right\}$$
$$-\frac{c}{4\pi}\left\{\left(\vec{\nabla}\cdot\left(\vec{E}_e+i\vec{E}_g\right)\right)\left(\vec{E}_e+i\vec{E}_g\right)+\left(\vec{\nabla}\cdot\left(\vec{H}_e+i\vec{H}_g\right)\right)\left(\vec{H}_e+i\vec{H}_g\right)\right\} \tag{5.8}$$
$$+c\left(\rho_e-i\rho_g\right)\left(\vec{E}_e+i\vec{E}_g\right)-i\left[\left(\vec{H}_e+i\vec{H}_g\right)\times\left(\vec{j}_e-i\vec{j}_g\right)\right]=0.$$

Finally, taking into account that $\varepsilon_1\varepsilon_2=0$ and separating the real and imaginary parts we get

$$\frac{1}{8\pi}\frac{\partial}{\partial t}\left\{\vec{E}_e^2+\vec{H}_e^2-\vec{E}_g^2-\vec{H}_g^2\right\}-i\frac{c}{4\pi}\left\{\left(\vec{\nabla}\cdot\left[\vec{E}_e\times\vec{H}_e\right]\right)-\left(\vec{\nabla}\cdot\left[\vec{E}_g\times\vec{H}_g\right]\right)\right\}$$
$$+\left\{\left(\vec{E}_e\cdot\vec{j}_e\right)+\left(\vec{E}_g\cdot\vec{j}_g\right)\right\}=0, \tag{5.9}$$

$$\frac{1}{8\pi}\vec{\nabla}\left\{\vec{E}_e^2+\vec{H}_e^2-\vec{E}_g^2-\vec{H}_g^2\right\}-i\frac{1}{4\pi c}\frac{\partial}{\partial t}\left\{\left[\vec{E}_e\times\vec{H}_e\right]-\left[\vec{E}_g\times\vec{H}_g\right]\right\}$$
$$-\frac{1}{4\pi}\left\{\left(\vec{\nabla}\cdot\vec{E}_e\right)\vec{E}_e+\left(\vec{\nabla}\cdot\vec{H}_e\right)\vec{H}_e-\left(\vec{\nabla}\cdot\vec{E}_g\right)\vec{E}_g-\left(\vec{\nabla}\cdot\vec{H}_g\right)\vec{H}_g\right\} \tag{5.10}$$
$$+\left\{\rho_e\vec{E}_e+\rho_g\vec{E}_g\right\}+i\left\{\left[\vec{H}_e\times\vec{j}_e\right]+\left[\vec{H}_g\times\vec{j}_g\right]\right\}=0,$$

$$\frac{1}{4\pi}\left\{\left(\vec{E}_e\cdot\frac{\partial\vec{H}_e}{\partial t}\right)-\left(\vec{H}_e\cdot\frac{\partial\vec{E}_e}{\partial t}\right)-\left(\vec{E}_g\cdot\frac{\partial\vec{H}_g}{\partial t}\right)+\left(\vec{H}_g\cdot\frac{\partial\vec{E}_g}{\partial t}\right)\right\}$$
$$-i\frac{c}{4\pi}\left\{\left(\vec{E}_e\cdot\left[\vec{\nabla}\times\vec{E}_e\right]\right)+\left(\vec{H}_e\cdot\left[\vec{\nabla}\times\vec{H}_e\right]\right)-\left(\vec{E}_g\cdot\left[\vec{\nabla}\times\vec{E}_g\right]\right)-\left(\vec{H}_g\cdot\left[\vec{\nabla}\times\vec{H}_g\right]\right)\right\} \tag{5.11}$$
$$-\left\{\left(\vec{H}_e\cdot\vec{j}_e\right)+\left(\vec{H}_g\cdot\vec{j}_g\right)\right\}=0,$$

$$-i\frac{1}{4\pi}\left\{\left[\vec{E}_e\times\frac{\partial\vec{E}_e}{\partial t}\right]-\left[\vec{E}_g\times\frac{\partial\vec{E}_g}{\partial t}\right]+\left[\vec{H}_e\times\frac{\partial\vec{H}_e}{\partial t}\right]-\left[\vec{H}_g\times\frac{\partial\vec{H}_g}{\partial t}\right]\right\}$$
$$+\frac{c}{4\pi}\left\{\vec{E}_e\left(\vec{\nabla}\cdot\vec{H}_e\right)-\vec{E}_g\left(\vec{\nabla}\cdot\vec{H}_g\right)-\vec{H}_e\left(\vec{\nabla}\cdot\vec{E}_e\right)+\vec{H}_g\left(\vec{\nabla}\cdot\vec{E}_g\right)\right\} \tag{5.12}$$
$$+\frac{c}{4\pi}\left\{\left[\vec{E}_e\times\left[\vec{\nabla}\times\vec{H}_e\right]\right]-\left[\vec{H}_e\times\left[\vec{\nabla}\times\vec{E}_e\right]\right]-\left[\vec{E}_g\times\left[\vec{\nabla}\times\vec{H}_g\right]\right]+\left[\vec{H}_g\times\left[\vec{\nabla}\times\vec{E}_g\right]\right]\right\}$$
$$+c\left\{\vec{H}_e\rho_e+\vec{H}_g\rho_g\right\}-i\left\{\left[\vec{E}_e\times\vec{j}_e\right]+\left[\vec{E}_g\times\vec{j}_g\right]\right\}=0.$$

The expression (5.9) is the generalized Pointing theorem for the GE field. The value $w$

$$w=\frac{1}{8\pi}\left\{\vec{E}_e^2+\vec{H}_e^2-\vec{E}_g^2-\vec{H}_g^2\right\} \tag{5.13}$$

plays the role of volume density of GE field energy, while vector $\vec{S}$

$$\vec{S}=-i\left\{\left[\vec{E}_e\times\vec{H}_e\right]-\left[\vec{E}_g\times\vec{H}_g\right]\right\} \tag{5.14}$$

plays the role of Pointing vector of GE field.

## 6. Lorentz invariants of GE field

The sedeonic algebra allows one to obtain relations for the Lorentz invariants of GE field. Let us multiply expression (4.22) on the complex conjugated sedeon of GE field intensities from the left:

$$\left(\mathbf{e}_{rt}\vec{E}_e+i\vec{H}_e-i\left(\mathbf{e}_{rt}\vec{E}_g+i\vec{H}_g\right)\right)\left(i\mathbf{e}_t\frac{1}{c}\frac{\partial}{\partial t}-\mathbf{e}_r\vec{\nabla}\right)\left(\mathbf{e}_{rt}\vec{E}_e-i\vec{H}_e+i\left(\mathbf{e}_{rt}\vec{E}_g-i\vec{H}_g\right)\right)$$
$$=-4\pi\left(\mathbf{e}_{rt}\vec{E}_e+i\vec{H}_e-i\left(\mathbf{e}_{rt}\vec{E}_g+i\vec{H}_g\right)\right)\left(i\mathbf{e}_t\rho_e+\mathbf{e}_r\frac{1}{c}\vec{j}_e-i\left(i\mathbf{e}_t\rho_g+\mathbf{e}_r\frac{1}{c}\vec{j}_g\right)\right). \tag{6.1}$$

Then performing sedeonic multiplication, we obtain the following expression:

$$-i\mathbf{e}_t \left\{ \frac{1}{c}\left((\vec{E}_e - i\vec{E}_g) \cdot \frac{\partial(\vec{E}_e + i\vec{E}_g)}{\partial t}\right) - \frac{1}{c}\left((\vec{H}_e - i\vec{H}_g) \cdot \frac{\partial(\vec{H}_e + i\vec{H}_g)}{\partial t}\right) \right.$$

$$\left. +i\left\{\left((\vec{E}_e - i\vec{E}_g) \cdot [\vec{\nabla} \times (\vec{H}_e + i\vec{H}_g)]\right) + \left((\vec{H}_e - i\vec{H}_g) \cdot [\vec{\nabla} \times (\vec{E}_e + i\vec{E}_g)]\right)\right\}\right\}$$

$$+i\mathbf{e}_r \frac{1}{c}\left\{\left((\vec{E}_e - i\vec{E}_g) \cdot \frac{\partial(\vec{H}_e + i\vec{H}_g)}{\partial t}\right) + \left((\vec{H}_e - i\vec{H}_g) \cdot \frac{\partial(\vec{E}_e + i\vec{E}_g)}{\partial t}\right)\right.$$

$$\left. -i\left((\vec{E}_e - i\vec{E}_g) \cdot [\vec{\nabla} \times (\vec{E}_e + i\vec{E}_g)]\right) + i\left((\vec{H}_e - i\vec{H}_g) \cdot [\vec{\nabla} \times (\vec{H}_e + i\vec{H}_g)]\right)\right\}$$

$$+\mathbf{e}_t \left\{ -i\frac{1}{c}\left[(\vec{E}_e - i\vec{E}_g) \times \frac{\partial(\vec{E}_e + i\vec{E}_g)}{\partial t}\right] + i\frac{1}{c}\left[(\vec{H}_e - i\vec{H}_g) \times \frac{\partial(\vec{H}_e + i\vec{H}_g)}{\partial t}\right]\right.$$

$$+(\vec{E}_e - i\vec{E}_g)(\vec{\nabla} \cdot (\vec{H}_e + i\vec{H}_g)) + (\vec{H}_e - i\vec{H}_g)(\vec{\nabla} \cdot (\vec{E}_e + i\vec{E}_g))$$

$$\left. +\left[(\vec{E}_e - i\vec{E}_g) \cdot [\vec{\nabla} \times (\vec{H}_e + i\vec{H}_g)]\right] + \left[(\vec{H}_e - i\vec{H}_g) \cdot [\vec{\nabla} \times (\vec{E}_e + i\vec{E}_g)]\right]\right\} \quad (6.2)$$

$$+\mathbf{e}_r \left\{ i\frac{1}{c}\left[(\vec{E}_e - i\vec{E}_g) \times \frac{\partial(\vec{H}_e + i\vec{H}_g)}{\partial t}\right] + i\frac{1}{c}\left[(\vec{H}_e - i\vec{H}_g) \times \frac{\partial(\vec{E}_e + i\vec{E}_g)}{\partial t}\right]\right.$$

$$+(\vec{E}_e - i\vec{E}_g)(\vec{\nabla} \cdot (\vec{E}_e + i\vec{E}_g)) - (\vec{H}_e - i\vec{H}_g)(\vec{\nabla} \cdot (\vec{H}_e + i\vec{H}_g))$$

$$\left. +\left[(\vec{E}_e - i\vec{E}_g) \times [\vec{\nabla} \times (\vec{E}_e + i\vec{E}_g)]\right] - \left[(\vec{H}_e - i\vec{H}_g) \times [\vec{\nabla} \times (\vec{H}_e + i\vec{H}_g)]\right]\right\}$$

$$= i\mathbf{e}_t \frac{4\pi}{c}\left((\vec{E}_e - i\vec{E}_g) \cdot (\vec{j}_e - i\vec{j}_g)\right) - i\mathbf{e}_r \frac{4\pi}{c}\left((\vec{H}_e - i\vec{H}_g) \cdot (\vec{j}_e - i\vec{j}_g)\right)$$

$$+4\pi\mathbf{e}_t \left\{(\rho_e - i\rho_g)(\vec{H}_e - i\vec{H}_g) + i\frac{1}{c}\left[(\vec{E}_e - i\vec{E}_g) \times (\vec{j}_e - i\vec{j}_g)\right]\right\}$$

$$+4\pi\mathbf{e}_r \left\{(\rho_e - i\rho_g)(\vec{E}_e - i\vec{E}_g) - i\frac{1}{c}\left[(\vec{H}_e - i\vec{H}_g) \times (\vec{j}_e - i\vec{j}_g)\right]\right\}.$$

Equating the components with different space-time properties, we get the following equations for GE field intensities:

$$\frac{1}{4\pi}\left\{\left((\vec{E}_e - i\vec{E}_g) \cdot \frac{\partial(\vec{E}_e + i\vec{E}_g)}{\partial t}\right) - \left((\vec{H}_e - i\vec{H}_g) \cdot \frac{\partial(\vec{H}_e + i\vec{H}_g)}{\partial t}\right)\right\}$$

$$+i\frac{c}{4\pi}\left\{\left((\vec{E}_e - i\vec{E}_g) \cdot [\vec{\nabla} \times (\vec{H}_e + i\vec{H}_g)]\right) + \left((\vec{H}_e - i\vec{H}_g) \cdot [\vec{\nabla} \times (\vec{E}_e + i\vec{E}_g)]\right)\right\} \quad (6.3)$$

$$+\left((\vec{E}_e - i\vec{E}_g) \cdot (\vec{j}_e - i\vec{j}_g)\right) = 0,$$

$$\frac{1}{4\pi}\left\{\left[\left(\vec{E}_e - i\vec{E}_g\right)\cdot\frac{\partial\left(\vec{H}_e + i\vec{H}_g\right)}{\partial t}\right] + \left(\left(\vec{H}_e - i\vec{H}_g\right)\cdot\frac{\partial\left(\vec{E}_e + i\vec{E}_g\right)}{\partial t}\right) - \right.$$
$$\left. -i\left(\left(\vec{E}_e - i\vec{E}_g\right)\cdot\left[\vec{\nabla}\times\left(\vec{E}_e + i\vec{E}_g\right)\right]\right) + i\left(\left(\vec{H}_e - i\vec{H}_g\right)\cdot\left[\vec{\nabla}\times\left(\vec{H}_e + i\vec{H}_g\right)\right]\right)\right\} \quad (6.4)$$
$$+\left(\left(\vec{H}_e - i\vec{H}_g\right)\cdot\left(\vec{j}_e - i\vec{j}_g\right)\right) = 0,$$

$$\frac{c}{4\pi}\left\{\left[\left(\vec{E}_e - i\vec{E}_g\right)\times\left[\vec{\nabla}\times\left(\vec{H}_e + i\vec{H}_g\right)\right]\right] + \left[\left(\vec{H}_e - i\vec{H}_g\right)\times\left[\vec{\nabla}\times\left(\vec{E}_e + i\vec{E}_g\right)\right]\right]\right\}$$
$$+\frac{c}{4\pi}\left\{\left(\vec{E}_e - i\vec{E}_g\right)\left(\vec{\nabla}\cdot\left(\vec{H}_e + i\vec{H}_g\right)\right) + \left(\vec{H}_e - i\vec{H}_g\right)\left(\vec{\nabla}\cdot\left(\vec{E}_e + i\vec{E}_g\right)\right)\right\}$$
$$-i\frac{1}{4\pi}\left\{\left[\left(\vec{E}_e - i\vec{E}_g\right)\times\frac{\partial\left(\vec{E}_e + i\vec{E}_g\right)}{\partial t}\right] - \left[\left(\vec{H}_e - i\vec{H}_g\right)\times\frac{\partial\left(\vec{H}_e + i\vec{H}_g\right)}{\partial t}\right]\right\} \quad (6.5)$$
$$-c\left(\vec{H}_e - i\vec{H}_g\right)\left(\rho_e - i\rho_g\right) - i\left[\left(\vec{E}_e - i\vec{E}_g\right)\times\left(\vec{j}_e - i\vec{j}_g\right)\right] = 0,$$

$$i\frac{1}{4\pi}\left\{\left[\left(\vec{E}_e - i\vec{E}_g\right)\times\frac{\partial\left(\vec{H}_e + i\vec{H}_g\right)}{\partial t}\right] + \left[\left(\vec{H}_e - i\vec{H}_g\right)\times\frac{\partial\left(\vec{E}_e + i\vec{E}_g\right)}{\partial t}\right]\right\}$$
$$+\frac{c}{4\pi}\left\{\left(\vec{E}_e - i\vec{E}_g\right)\left(\vec{\nabla}\cdot\left(\vec{E}_e + i\vec{E}_g\right)\right) - \left(\vec{H}_e - i\vec{H}_g\right)\left(\vec{\nabla}\cdot\left(\vec{H}_e + i\vec{H}_g\right)\right)\right\} \quad (6.6)$$
$$+\frac{c}{4\pi}\left\{\left[\left(\vec{E}_e - i\vec{E}_g\right)\times\left[\vec{\nabla}\times\left(\vec{E}_e + i\vec{E}_g\right)\right]\right] - \left[\left(\vec{H}_e - i\vec{H}_g\right)\times\left[\vec{\nabla}\times\left(\vec{H}_e + i\vec{H}_g\right)\right]\right]\right\}$$
$$-c\left(\rho_e - i\rho_g\right)\left(\vec{E}_e - i\vec{E}_g\right) - i\left[\left(\vec{H}_e - i\vec{H}_g\right)\times\left(\vec{j}_e - i\vec{j}_g\right)\right] = 0.$$

Finally, taking into account that $\varepsilon_1\varepsilon_2 = 0$ and separating the real and imaginary parts we get

$$\frac{1}{8\pi}\frac{\partial}{\partial t}\left\{\vec{E}_e^2 - \vec{H}_e^2 + \vec{E}_g^2 - \vec{H}_g^2\right\}$$
$$+i\frac{c}{4\pi}\left\{\left(\vec{E}_e\cdot\left[\vec{\nabla}\times\vec{H}_e\right]\right) + \left(\vec{H}_e\cdot\left[\vec{\nabla}\times\vec{E}_e\right]\right) + \left(\vec{E}_g\cdot\left[\vec{\nabla}\times\vec{H}_g\right]\right) + \left(\vec{H}_g\cdot\left[\vec{\nabla}\times\vec{E}_g\right]\right)\right\} \quad (6.7)$$
$$+\left\{\left(\vec{E}_e\cdot\vec{j}_e\right) - \left(\vec{E}_g\cdot\vec{j}_g\right)\right\} = 0,$$

$$\frac{c}{8\pi}\vec{\nabla}\left\{\vec{E}_e^2 - \vec{H}_e^2 + \vec{E}_g^2 - \vec{H}_g^2\right\}$$
$$-i\frac{1}{4\pi}\left\{\left[\vec{E}_e\times\frac{\partial\vec{H}_e}{\partial t}\right] + \left[\vec{E}_g\times\frac{\partial\vec{H}_g}{\partial t}\right] + \left[\vec{H}_e\times\frac{\partial\vec{E}_e}{\partial t}\right] + \left[\vec{H}_g\times\frac{\partial\vec{E}_g}{\partial t}\right]\right\}$$
$$-\frac{c}{4\pi}\left\{\vec{E}_e\left(\vec{\nabla}\cdot\vec{E}_e\right) + \left(\vec{E}_e\cdot\vec{\nabla}\right)\vec{E}_e + \vec{E}_g\left(\vec{\nabla}\cdot\vec{E}_g\right) + \left(\vec{E}_g\cdot\vec{\nabla}\right)\vec{E}_g\right. \quad (6.8)$$
$$\left. -\vec{H}_e\left(\vec{\nabla}\cdot\vec{H}_e\right) - \left(\vec{H}_e\cdot\vec{\nabla}\right)\vec{H}_e - \vec{H}_g\left(\vec{\nabla}\cdot\vec{H}_g\right) - \left(\vec{H}_g\cdot\vec{\nabla}\right)\vec{H}_g\right\}$$
$$+c\left\{\rho_e\vec{E}_e - \rho_g\vec{E}_g\right\} - i\left\{\left[\vec{H}_e\times\vec{j}_e\right] - \left[\vec{H}_g\times\vec{j}_g\right]\right\} = 0,$$

$$\frac{1}{4\pi}\frac{\partial}{\partial t}\left\{\left(\vec{E}_e\cdot\vec{H}_e\right)+\left(\vec{E}_g\cdot\vec{H}_g\right)\right\}$$

$$-i\frac{c}{4\pi}\left\{\left(\vec{E}_e\cdot\left[\vec{\nabla}\times\vec{E}_e\right]\right)-\left(\vec{H}_e\cdot\left[\vec{\nabla}\times\vec{H}_e\right]\right)+\left(\vec{E}_g\cdot\left[\vec{\nabla}\times\vec{E}_g\right]\right)-\left(\vec{H}_g\cdot\left[\vec{\nabla}\times\vec{H}_g\right]\right)\right\} \quad (6.9)$$

$$+\left\{\left(\vec{H}_e\cdot\vec{j}_e\right)-\left(\vec{H}_g\cdot\vec{j}_g\right)\right\}=0,$$

$$\frac{c}{4\pi}\vec{\nabla}\left\{\left(\vec{E}_e\cdot\vec{H}_e\right)+\left(\vec{E}_g\cdot\vec{H}_g\right)\right\}$$

$$-\frac{c}{4\pi}\left\{\vec{E}_e\left(\vec{\nabla}\cdot\vec{H}_e\right)+\vec{E}_g\left(\vec{\nabla}\cdot\vec{H}_g\right)+\vec{H}_e\left(\vec{\nabla}\cdot\vec{E}_e\right)+\vec{H}_g\left(\vec{\nabla}\cdot\vec{E}_g\right)\right.$$

$$\left.-\left(\vec{E}_e\cdot\vec{\nabla}\right)\vec{H}_e+\left(\vec{E}_g\cdot\vec{\nabla}\right)\vec{H}_g+\left(\vec{H}_e\cdot\vec{\nabla}\right)\vec{E}_e+\left(\vec{H}_g\cdot\vec{\nabla}\right)\vec{E}_g\right\} \quad (6.10)$$

$$+i\frac{1}{4\pi}\left\{\left[\vec{E}_e\times\frac{\partial\vec{E}_e}{\partial t}\right]+\left[\vec{E}_g\times\frac{\partial\vec{E}_g}{\partial t}\right]-\left[\vec{H}_e\times\frac{\partial\vec{H}_e}{\partial t}\right]-\left[\vec{H}_g\times\frac{\partial\vec{H}_g}{\partial t}\right]\right\}$$

$$+c\left\{\vec{H}_e\rho_e-\vec{H}_g\rho_g\right\}-i\left\{\left[\vec{E}_e\times\vec{j}_e\right]-\left[\vec{E}_g\times\vec{j}_g\right]\right\}=0.$$

The expressions (6.7)-(6.10) are the equations for the generalized Lorentz invariants $I_1$ and $I_2$ of GE field:

$$I_1 = \vec{E}_e^2 - \vec{H}_e^2 + \vec{E}_g^2 - \vec{H}_g^2, \quad (6.11)$$

$$I_2 = \left(\vec{E}_e\cdot\vec{H}_e\right)+\left(\vec{E}_g\cdot\vec{H}_g\right). \quad (6.12)$$

## 7. Sedeonic equations for neutrino field

The free massless neutrino field is described by the first-order sedeonic equation [27,28]:

$$\left(i\mathbf{e_t}\frac{1}{c}\frac{\partial}{\partial t}-\mathbf{e_r}\vec{\nabla}\right)\tilde{W}_\nu = 0. \quad (7.1)$$

Drawing on an analogy with the gravitoelectromagnetic field (see Eq. 4.2) we can write the potential $\tilde{W}_\nu$ in the following form:

$$\tilde{W}_\nu = i\mathbf{e_t}\varphi_\nu + \mathbf{e_r}\vec{A}_\nu, \quad (7.2)$$

where $\varphi_\nu$ and $\vec{A}_\nu$ are complex scalar and vector potentials of neutrino field:

$$\varphi_\nu = \varphi_\alpha + i\varphi_\beta, \quad (7.3)$$

$$\vec{A}_\nu = \vec{A}_\alpha + i\vec{A}_\beta. \quad (7.4)$$

Thus the equation for free neutrino field can be represented as

$$\left(i\mathbf{e_t}\frac{1}{c}\frac{\partial}{\partial t}-\mathbf{e_r}\vec{\nabla}\right)\left(i\mathbf{e_t}\varphi_\nu + \mathbf{e_r}\vec{A}_\nu\right) = 0. \quad (7.5)$$

Applying the operator

$$i\mathbf{e_t}\frac{1}{c}\frac{\partial}{\partial t}-\mathbf{e_r}\vec{\nabla}$$

to the equation (7.5) we get

$$\left(\frac{1}{c^2}\frac{\partial^2}{\partial t^2}-\Delta\right)\left(i\mathbf{e}_t\varphi_v+\mathbf{e}_r\vec{A}_v\right)=0. \tag{7.6}$$

Separating the real and imaginary terms with different space-time properties, we get the following wave equations for the potentials:

$$\left(\frac{1}{c^2}\frac{\partial^2}{\partial t^2}-\Delta\right)\varphi_\alpha=0, \tag{7.7}$$

$$\left(\frac{1}{c^2}\frac{\partial^2}{\partial t^2}-\Delta\right)\varphi_\beta=0, \tag{7.8}$$

$$\left(\frac{1}{c^2}\frac{\partial^2}{\partial t^2}-\Delta\right)\vec{A}_\alpha=0. \tag{7.9}$$

$$\left(\frac{1}{c^2}\frac{\partial^2}{\partial t^2}-\Delta\right)\vec{A}_\beta=0. \tag{7.10}$$

In fact, the potentials of neutrino field $\varphi_\alpha$, $\varphi_\beta$, $\vec{A}_\alpha$, $\vec{A}_\beta$ satisfy the second order wave equations analogues to the equations for GE field. However, first order equation (7.5) describes the massless field with field intensities equal to zero (see the expression 4.11 for comparison) [27,30].

On the other hand performing sedeonic multiplication in equation (7.5) we get

$$-\frac{1}{c}\frac{\partial\varphi_v}{\partial t}-\mathbf{e}_{tr}\frac{1}{c}\frac{\partial\vec{A}_v}{\partial t}-\mathbf{e}_{tr}\vec{\nabla}\varphi_v-\left(\vec{\nabla}\cdot\vec{A}_v\right)-\left[\vec{\nabla}\times\vec{A}_v\right]=0. \tag{7.12}$$

Separating the terms with different space-time properties, we get the following system of equations for the neutrino field potentials:

$$\begin{aligned}&\frac{1}{c}\frac{\partial\varphi_\alpha}{\partial t}+\left(\vec{\nabla}\cdot\vec{A}_\alpha\right)=0,\\ &\frac{1}{c}\frac{\partial\vec{A}_\alpha}{\partial t}+\vec{\nabla}\varphi_\alpha=0,\\ &\left[\vec{\nabla}\times\vec{A}_\alpha\right]=0,\\ &\frac{1}{c}\frac{\partial\varphi_\beta}{\partial t}+\left(\vec{\nabla}\cdot\vec{A}_\beta\right)=0,\\ &\frac{1}{c}\frac{\partial\vec{A}_\beta}{\partial t}+\vec{\nabla}\varphi_\beta=0,\\ &\left[\vec{\nabla}\times\vec{A}_\beta\right]=0.\end{aligned} \tag{7.13}$$

Thus, based on analogy with the electromagnetic and gravitational fields (comaring equations (4.4) and (7.6)) one can assume that the generalized equation (7.5) describes the special field of a gravitoelectromagnetic nature. The potentials $\varphi_\alpha$ and $\vec{A}_\alpha$ describe the electromagnetic component, while the potentials $\varphi_\beta$ and $\vec{A}_\beta$ describe the gravitational component of the neutrino field. Below we will assume that potentials $\varphi_\alpha$ and $\vec{A}_\alpha$ contain $\varepsilon_1$ unit, while $\varphi_\beta$ and $\vec{A}_\beta$ contain $\varepsilon_2$ unit.

## 8. Second-order relations for neutrino field

Multiplying the expression (7.5) on potential $\tilde{W}_\nu$ from the left, we obtain the following sedeonic equation:

$$\left(i\mathbf{e_t}\varphi_\nu + \mathbf{e_r}\vec{A}_\nu\right)\left(i\mathbf{e_t}\frac{1}{c}\frac{\partial}{\partial t} - \mathbf{e_r}\vec{\nabla}\right)\left(i\mathbf{e_t}\varphi_\nu + \mathbf{e_r}\vec{A}_\nu\right) = 0. \tag{8.1}$$

Performing the sedeonic multiplication and separating different terms we get second order expressions for the neutrino field potentials:

$$\frac{1}{2c}\frac{\partial}{\partial t}\left\{\varphi_\nu^2 + \vec{A}_\nu^2\right\} + \left(\vec{\nabla}\cdot\varphi_\nu\vec{A}_\nu\right) = 0, \tag{8.2}$$

$$\left(\vec{A}_\nu\cdot\left[\vec{\nabla}\times\vec{A}_\nu\right]\right) = 0, \tag{8.3}$$

$$\frac{1}{c}\left[\vec{A}_\nu\times\frac{\partial\vec{A}_\nu}{\partial t}\right] + \left[\varphi_\nu\vec{\nabla}\times\vec{A}_\nu\right] + \left[\vec{A}_\nu\times\vec{\nabla}\varphi_\nu\right] = 0, \tag{8.4}$$

$$\frac{1}{c}\frac{\partial}{\partial t}\left\{\varphi_\nu\vec{A}_\nu\right\} + \frac{1}{2}\vec{\nabla}\left\{\varphi_\nu^2 - \vec{A}_\nu^2\right\} + \left(\vec{\nabla}\cdot\vec{A}_\nu\right)\vec{A}_\nu = 0. \tag{8.5}$$

Separating the real and imaginary parts and excluding the cross-terms (taking into account that $\varepsilon_1\varepsilon_2 = 0$) we get following four equations:

$$\frac{1}{2c}\frac{\partial}{\partial t}\left\{\varphi_\alpha^2 + \vec{A}_\alpha^2 - \varphi_\beta^2 - \vec{A}_\beta^2\right\} + \left\{\left(\vec{\nabla}\cdot\varphi_\alpha\vec{A}_\alpha\right) - \left(\vec{\nabla}\cdot\varphi_\beta\vec{A}_\beta\right)\right\} = 0, \tag{8.6}$$

$$\frac{1}{2}\vec{\nabla}\left\{\varphi_\alpha^2 - \vec{A}_\alpha^2 - \varphi_\beta^2 + \vec{A}_\beta^2\right\} + \frac{1}{c}\frac{\partial}{\partial t}\left\{\varphi_\alpha\vec{A}_\alpha - \varphi_\beta\vec{A}_\beta\right\}$$
$$+ \left\{\left(\vec{\nabla}\cdot\vec{A}_\alpha\right)\vec{A}_\alpha - \left(\vec{\nabla}\cdot\vec{A}_\beta\right)\vec{A}_\beta\right\} = 0, \tag{8.7}$$

$$\left(\vec{A}_\alpha\cdot\left[\vec{\nabla}\times\vec{A}_\alpha\right]\right) - \left(\vec{A}_\beta\cdot\left[\vec{\nabla}\times\vec{A}_\beta\right]\right) = 0, \tag{8.8}$$

$$\frac{1}{c}\left\{\left[\vec{A}_\alpha\times\frac{\partial\vec{A}_\alpha}{\partial t}\right] - \left[\vec{A}_\beta\times\frac{\partial\vec{A}_\beta}{\partial t}\right]\right\}$$
$$+ \left\{\left[\varphi_\alpha\vec{\nabla}\times\vec{A}_\alpha\right] - \left[\varphi_\beta\vec{\nabla}\times\vec{A}_\beta\right] + \left[\vec{A}_\alpha\times\vec{\nabla}\varphi_\alpha\right] - \left[\vec{A}_\beta\times\vec{\nabla}\varphi_\beta\right]\right\} = 0. \tag{8.9}$$

On the other hand, multiplying the expression (7.5) on $\left(-i\mathbf{e_t}\varphi_\nu + \mathbf{e_r}\vec{A}_\nu\right)$ from the left, we obtain the following sedeonic equation:

$$\left(-i\mathbf{e_t}\varphi_\nu + \mathbf{e_r}\vec{A}_\nu\right)\left(i\mathbf{e_t}\frac{1}{c}\frac{\partial}{\partial t} - \mathbf{e_r}\vec{\nabla}\right)\left(i\mathbf{e_t}\varphi_\nu + \mathbf{e_r}\vec{A}_\nu\right) = 0. \tag{8.10}$$

Performing the sedeonic multiplication and separating different terms we get following expressions

$$\frac{1}{2c}\frac{\partial}{\partial t}\left\{\varphi_\nu^2 - \vec{A}_\nu^2\right\} + \varphi_\nu\left(\vec{\nabla}\cdot\vec{A}_\nu\right) - \left(\vec{A}_\nu\cdot\vec{\nabla}\right)\varphi_\nu = 0, \tag{8.11}$$

$$\left(\vec{A}_\nu\cdot\left[\vec{\nabla}\times\vec{A}_\nu\right]\right) = 0, \tag{8.12}$$

$$\frac{1}{c}\left[\vec{A}_\nu \times \frac{\partial \vec{A}_\nu}{\partial t}\right] - \left[\vec{\nabla} \times \varphi_\nu \vec{A}_\nu\right] = 0, \tag{8.13}$$

$$\frac{1}{c}\left\{\varphi_\nu \frac{\partial \vec{A}_\nu}{\partial t} - \vec{A}_\nu \frac{\partial \varphi_\nu}{\partial t}\right\} + \frac{1}{2}\vec{\nabla}\left\{\varphi_\nu^2 + \vec{A}_\nu^2\right\} - \left(\vec{\nabla} \cdot \vec{A}_\nu\right)\vec{A}_\nu = 0. \tag{8.14}$$

Separating the real and imaginary parts and excluding the cross-terms we get another four equations:

$$\begin{aligned}&\frac{1}{2c}\frac{\partial}{\partial t}\left\{\varphi_\alpha^2 - \vec{A}_\alpha^2 - \varphi_\beta^2 + \vec{A}_\beta^2\right\} \\ &+ \left\{\varphi_\alpha\left(\vec{\nabla} \cdot \vec{A}_\alpha\right) + \varphi_\beta\left(\vec{\nabla} \cdot \vec{A}_\beta\right) - \left(\vec{A}_\alpha \cdot \vec{\nabla}\right)\varphi_\alpha - \left(\vec{A}_\beta \cdot \vec{\nabla}\right)\varphi_\beta\right\} = 0,\end{aligned} \tag{8.15}$$

$$\begin{aligned}&\frac{1}{2}\vec{\nabla}\left\{\varphi_\alpha^2 + \vec{A}_\alpha^2 - \varphi_\beta^2 - \vec{A}_\beta^2\right\} + \frac{1}{c}\left\{\varphi_\alpha \frac{\partial \vec{A}_\alpha}{\partial t} - \vec{A}_\alpha \frac{\partial \varphi_\alpha}{\partial t} + \vec{A}_\beta \frac{\partial \varphi_\beta}{\partial t} - \varphi_\beta \frac{\partial \vec{A}_\beta}{\partial t}\right\} \\ &- \left\{\left(\vec{\nabla} \cdot \vec{A}_\alpha\right)\vec{A}_\alpha + \left(\vec{\nabla} \cdot \vec{A}_\beta\right)\vec{A}_\beta\right\} = 0,\end{aligned} \tag{8.16}$$

$$\left(\vec{A}_\alpha \cdot \left[\vec{\nabla} \times \vec{A}_\alpha\right]\right) - \left(\vec{A}_\beta \cdot \left[\vec{\nabla} \times \vec{A}_\beta\right]\right) = 0, \tag{8.17}$$

$$\frac{1}{c}\left\{\left[\vec{A}_\alpha \times \frac{\partial \vec{A}_\alpha}{\partial t}\right] - \left[\vec{A}_\beta \times \frac{\partial \vec{A}_\beta}{\partial t}\right]\right\} - \left\{\left[\vec{\nabla} \times \varphi_\alpha \vec{A}_\beta\right] - \left[\vec{\nabla} \times \varphi_\beta \vec{A}_\alpha\right]\right\} = 0. \tag{8.18}$$

The expressions (8.6), (8.7), (8.15) and (8.16) are the analogs of Poynting theorem and Lorentz invariants relations for the neutrino field.

## 9. Summary

Thus in this paper we have developed the description of massless fields on the basis of space-time algebra of sixteen-component sedeons. The generalized sedeonic second-order equation for unified gravitoelectromagnetic field describing simultaneously gravity and electromagnetism was proposed. We have derived the relations for energy, momentum and Lorentz invariants of unified GE field. Besides, we considered the generalized sedeonic first-order equation for the neutrino field. Using the analogy with the GE field one can assume that this equation describes the electromagnetic and gravitational components of the neutrino field. The second-order relations for the neutrino potentials (analogs of Pointing theorem and Lorentz invariants relations for GE field) were also derived.

**Acknowledgments**

The authors are very thankful to G.V. Mironova for kind assistance and moral support.

**References**

1. S.Weinberg, *Gravitation and cosmology*, (Jon Wiley, New York, 1972).
2. L.D.Landau, E.M.Lifschitz, *The classical theory of fields*, (Pergamon Press, Oxford, 1987).
3. B.Mashhoon, Gravitoelectromagnetism: A brief review, in *The Measurement of Gravitomagnetism: A Challenging Enterprise*, edited by L.Iorio (NOVA Science, Hauppage, New York, 2007) ch. 3, pp. 29-39.
4. J.D.Kaplan, D.A.Nichols, K.S.Thorne, Post-Newtonian approximation in Maxwell-like form, Physical Review D, **80**, 124014 1-6 (2009).

5. S.M.Kopeikin, Gravitomagnetism and the speed of gravity, International Journal of Modern Physics D, **15**, 305-320 (2006).
6. M.L.Ruggiero, A.Tartaglia, Gravitomagnetic effects, Il Nuovo Cimento B, **117**, 743-768 (2002).
7. V.Majernik, Field approach to gravitation and its significance in astrophysics, Astrophysics and Space Science, **14**, 265-285 (1971).
8. L.Iorio, Recent developments in testing gravitomagnetism with satellite laser ranging, in *The Measurement of Gravitomagnetism: A Challenging Enterprise,* edited by L.Iorio (NOVA Science, Hauppage, New York, 2007) ch. 8, pp. 103-136.
9. G.Schäfer, Gravitomagnetism in physics and astrophysics, Space Science Review, **148**, 37-52 (2009).
10. T.Damour, M.Soffel, Ch.Xu, General-relativistic celestial mechanics. I. Method and definition reference systems, Physical Review D, **43**, 3273-3307 (1991).
11. V.Majernik, Quaternionic formulation of the classical fields, Advances in Applied Clifford Algebras, **9**(1), 119-130 (1999).
12. K.Imaeda, A new formulation of classical electrodynamics, Il Nuovo cimento B, **32**(1) 138-162 (1976).
13. M.Gogberashvili, Octonionic electrodynamics, Journal of Physics A.: Mathematical and General, **39**, 7099-7104, (2006).
14. A.Gamba, Maxwell's equations in octonion form, Il Nuovo Cimento A, **111**(3), 293-302, (1998).
15. T.Tolan, K.Özdas, M.Tanisli, Reformulation of electromagnetism with octonions, Il Nuovo Cimento B, **121**(1), 43-55, (2006).
16. A.M.Shaarawi, Clifford algebra formulation of an electromagnetic charge-current wave theory, Foundations of physics, **30**(11), 1911-1941, (2000).
17. V.L.Mironov, S.V.Mironov, Octonic representation of electromagnetic field equations, Journal of Mathematical Physics, **50**, 012901 (2009).
18. V.Majernik, Some astrophysical consequences of the extended Maxwell-like gravitational field equations, Astrophysics and Space Science, **84**, 191-204 (1982).
19. S.Ulrych, Gravitoelectromagnetism in a complex Clifford algebra, Physics Letters B, **633**, 631-635 (2006).
20. K.Imaeda, M.Imaeda, Sedenions: algebra and analysis, Appl. Math. Comp., **115**, 77-88 (2000).
21. K.Carmody, Circular and hyperbolic quaternions, octonions, and sedenions, Applied Mathematics and Computation, **28**, 47-72 (1988).
22. K.Carmody, Circular and hyperbolic quaternions, octonions, and sedenions – further results, Applied Mathematics and Computation, **84**, 27-47 (1997).
23. J.Köplinger, Dirac equation on hyperbolic octonions, Applied Mathematics and Computation, **182**, 443-446 (2006).
24. S.Demir, M.Tanisli, Sedenionic formulation for generalized fields of dyons, International Journal of Theoretical Physics, **51**(4), 1239-1253 (2012).
25. W.P.Joyce, Dirac theory in spacetime algebra: I. The generalized bivector Dirac equation, Journal of Physics A: Mathematical and General, **34**, 1991-2005 (2001).
26. C.Cafaro, S.A. Ali, The spacetime algebra approach to massive classical electrodynamics with magnetic monopoles, Advances in Applied Clifford Algebras, **17**, 23-36 (2006).
27. V.L.Mironov, S.V.Mironov, Sedeonic generalization of relativistic quantum mechanics, International Journal of Modern Physics A, **24**(32), 6237-6254 (2009).
28. V.L.Mironov, S.V.Mironov, Noncommutative sedeons and their application in field theory, e-print: arXiv:1111.4035v1 [math-ph], (2011).
29. S.Schwebel, Newtonian gravitational field theory, International Journal of Theoretical Physics, **3**(4), 315-330 (1970).
30. V.L.Mironov, S.V.Mironov, Octonic first-order equations of relativistic quantum mechanics, International Journal of Modern Physics A, **24**(22), 4157-4167 (2009).